\documentclass{SCGE}

\begin{document}

%%%%%%аæʽҪ¼ÓÉÏÕâ×é
\begin{picture}(0,0){\rm
\put(0,-20){\makebox[160truemm][l]{\bf {\sanhao\raisebox{2pt}{.}}
Article  {\sanhao\raisebox{1.5pt}{.}}}}}
\put(0,-34){\jiuwuhao {\textcolor[rgb]{0.5,0.5,0.5}{\sf %Special Topic: Fluid Mechanics
}}}%%(11ÔÂ×¢ÊÍ£ºµ÷\textcolor[rgb]{x,x,x}ÖеÄÊý×ÖxÔ½´óÔ½»Ò)
\end{picture}

\def\bm{\boldsymbol}

\def\dl{\displaystyle}
\def\du{\end{document}}
\def\d{{\rm d}}
\def\e{{\rm e}}
\def\i{{\rm i}}
\def\pi{{\uppi}}

% The author doesn't need fill in it.
\Year{2016} %
\Month{??} %
\Vol{??} %  ¾íºÅ
\No{?} %  ÆÚºÅ
\BeginPage{1} % ÆðÒ³Âë
\AuthorMark{{\rm Zhou L J}, et al.}  %(11ÔÂ×¢ÊÍ£ºÒ³Ã¼ÉϵÄ×÷Õß)
\AuthorMarkCite{{\rm Zhou L J, \& Wang, S} } %(11ÔÂ×¢ÊÍ£ºcitationÖеÄ×÷Õß)
\DOI{??} % The author doesn't need fill in it.
\ArtNo{??}

% \title[short text for running head]{full title}{comments for title}
\title[Diagnosing $\Lambda$HDE model with statefinder hierarchy and fractional growth parameter]
{Diagnosing $\Lambda$HDE model with statefinder hierarchy and fractional growth parameter}

\author[1]{ZHOU Lanjun}{}
\author[2]{WANG Shuang}{}

\address[{\rm1}]{Institute of Theoretical Physics, Chinese Academy of Sciences, Beijing 100190, P. R. China;}
\address[{\rm2}]{School of Astronomy and Space Science, Sun Yat-Sen University, Guangzhou 510275, P. R. China}

\maketitle \vspace{-3.5mm}{\footnotesize\begin{center} %Received January 23, 2015; accepted February 3, 2015; published online March 5, 2015
\end{center}}\vspace*{-5mm}

%     Abstract is required.
\begin{center}
\rule{16.5cm}{0.4pt}
\parbox{16.5cm}
{\begin{abstract}
Recently, a new dark energy model called $\Lambda$HDE was proposed.
In this model, dark energy consists of two parts: cosmological constant $\Lambda$ and holographic dark energy (HDE).
Two key parameters of this model are the fractional density of cosmological constant $\Omega_{\Lambda0}$, and the dimensionless HDE parameter $c$.
Since these two parameters determine the dynamical properties of DE and the destiny of universe, it is important to study the impacts of different values of $\Omega_{\Lambda0}$ and $c$ on the $\Lambda$HDE model.
In this paper, we apply various DE diagnostic tools to diagnose $\Lambda$HDE models with different values of $\Omega_{\Lambda0}$ and $c$;
these tools include statefinder hierarchy \{$S_3^{(1)}, S_4^{(1)}$\}, fractional growth parameter $\epsilon$, and composite null diagnostic (CND), which is a combination of \{$S_3^{(1)}, S_4^{(1)}$\} and $\epsilon$.
We find that: (1) adopting different values of $\Omega_{\Lambda0}$ only has quantitative impacts on the evolution of the $\Lambda$HDE model, while adopting different $c$ has qualitative impacts;
(2) compared with $S_3^{(1)}$, $S_4^{(1)}$ can give larger differences among the cosmic evolutions of the $\Lambda$HDE model associated with different $\Omega_{\Lambda0}$ or different $c$;
(3) compared with the case of using a single diagnostic, adopting a CND pair has much stronger ability to diagnose the $\Lambda$HDE model.
\end{abstract}}
\end{center}\vspace*{-0.6cm}

\begin{center}
\parbox{16.5cm}
{\bf\jiuhao Key words: Dark Energy, Cosmology, Cosmological Constant}
\end{center}

\begin{center}
{}%·ÖÀàºÅ
\CITA    %%(11ÔÂ×¢ÊÍ£ºCitationÄÚÈÝ×Ô¶¯Éú³É)
%\Cit{~~~???, et al. ???. Sci China-Phys Mech Astron, 2014, 57: 1--6, doi:}%%(11ÔÂ×¢ÊÍ£ºCitationÄÚÈÝÐèÊÖ¶¯Ìîд)
\end{center}

\textwidth=178truemm \textheight=236truemm%%%%%%аæʽҪ¼ÓÉÏ

%%%%%%%%%%%%%%%%%%%%%%%%%%%%%%%%%%%%%%%%%%%%%%%%%%%%%%%%%%%%
\wuhao\vspace*{1.5mm}

\begin{multicols}{2}

%%%%%%%%%%%%%%%%%%%%%%%%%%%%%%%%%%%%%%%%%%%%%%%%%%%%%%%%%%%%
%% Text of article.
%%%%%%%%%%%%%%%%%%%%%%%%%%%%%%%%%%%%%%%%%%%%%%%%%%%%%%%%%%%%
%    Section headings
\renewcommand{\baselinestretch}{1.08} \baselineskip 12.2pt\parindent=10.8pt

\renewcommand{\thefootnote}

\section{Introduction}\label{intro}

Dark energy (DE) has become one of the most important problems in modern cosmology \cite{ref_observ1,ref_observ2,ref_derev1,ref_derev2}.
Although numerous DE models \cite{ref_quinte,ref_phantom,ref_k,ref_Chaplygin,ref_YMC,ref_IYMC,ref_bin1,ref_IDEage} have been proposed,
the nature of DE is still in dark.

In principle, the DE problem may be an issue of quantum gravity \cite{ref_Witten}.
It is commonly believed that the holographic principle \cite{ref_Hooft} is just a fundamental principle of quantum gravity.
Based on holographic principle,
Li \cite{ref_hde} firstly proposed a promising DE model,
which is called holographic dark energy (HDE) model.
In this model, the density of HDE can be written as

\noindent\rule{2.5cm}{0.4pt}\\[0.1mm]{\qihao ljzhou@itp.ac.cn\vspace*{-1.3mm}\\
wangshuang@mail.sysu.edu.cn (Corresponding author)}%ÊÖ¶¯E-mailµØÖ·

\begin{equation}
\rho_{hde}=3c^2 M_p^2L^{-2},
\end{equation}
where $c$ is a dimensionless parameter and $M_p$ is the reduced Planck mass.
$L$ is the IR cutoff length scale, which takes the form \cite{ref_hde_new}:
\begin{equation}
  L=\frac{a}{\sqrt{|k|}} sinn(\sqrt{|k|} \int_t^{+\infty} \frac{dt'}{a}),
\end{equation}
where $a$ is the scale factor, $k$ is a constant representing the space curvature, and the function $sinn(x)$ is defined as
\begin{eqnarray}
  sinn(x) &=& \left\{ \begin{array}{ll}
  sin(x), & \textrm{if $k>0$}\\
  x,      & \textrm{if $k=0$}\\
  sinh(x),& \textrm{if $k<0$}
\end{array} \right.
\end{eqnarray}
The HDE model is the first theoretical model inspired by holographic principle;
in addition, it is in good agreement with the current cosmological observations.
Therefore, in recent years, this model has drawn a lot of attention and has been widely studied in the literature \cite{ref_hde1, ref_hde2, ref_hde3, ref_hde4, ref_hde5, ref_hde6, ref_hde7, ref_hde8}.

In a latest paper \cite{ref_lhde}, inspired by the multiverse scenario, a new model called $\Lambda$HDE was proposed.
In this model, dark energy consists of two parts: cosmological constant $\Lambda$ and HDE.
Now the density of total DE is
\begin{equation}
\rho_{de} = \rho_\Lambda+\rho_{hde},
\end{equation}
Both the theoretical implications and observational constraints were simply discussed in \cite{ref_lhde}.

It should be mentioned that the $\Lambda$HDE model has two key parameters: fractional density of cosmological constant $\Omega_{\Lambda0}$ and dimensionless HDE parameter $c$.
Since they determine the dynamical properties of DE and the destiny of universe, it is important to study the impacts of different values of $\Omega_{\Lambda0}$ and $c$ on the $\Lambda$HDE model.
Two diagnostic tools are often used to analyze various DE models.
The first one is statefinder hierarchy \cite{ref_sttfd1, ref_sttfd2},
which is a model-independent geometrical diagnostic tool.
The second one is the fractional growth parameter $\epsilon$ \cite{ref_growth1, ref_growth2},
which provides a scale-independent diagnostic of growth history of universe.
In addition, a combination of statefinder hierarchy and fractional growth parameter, which is called composite null diagnostic (CND) \cite{ref_sttfd2}, is often used to diagnose DE models.
The main aim of this work is making use of these tools to distinguish the $\Lambda$HDE models with different $\Omega_{\Lambda0}$ or different $c$.

This paper is organized as follows. In Sec. \ref{model}, we briefly review the $\Lambda$HDE model. In Sec. \ref{tools}, we introduce the diagnostic tools used in this work. In Sec. \ref{result}, we present the obtained results. Conclusions and discussions are given in Sec. \ref{summary}.
\vspace*{-1mm}

\section{Theoretical model} \label{model} \vspace*{-1mm}

In this section, we briefly introduce how to calculate the evolution of reduced Hubble parameter $E(z)\equiv H/H_0$ for the $\Lambda$HDE model, where $H$ and $H_0$ denote Hubble parameter and its present-day value.
We will neglect the effect of radiation component, because in this paper we focus on dark energy which has effect only in low-redshift region.
Thus, $E(z)$ and fractional HDE energy density $\Omega_{hde}$ for $\Lambda$HDE model are determined by the following equations \cite{ref_lhde}:

\begin{equation}
  \label{elhde}
  \small
  {1\over E(z)}{dE(z) \over dz} = -{\Omega_{hde}\over
  1+z}\left({3\Omega_{\Lambda}+\Omega_k-3\over2\Omega_{hde}}+{1\over2}+\sqrt{{\Omega_{hde}\over c^2}+\Omega_k} \right),
\end{equation}
\begin{equation}
  \label{Ohde}
  \footnotesize
    {d\Omega_{hde}\over dz} =
  -{2\Omega_{hde}(1-\Omega_{hde})\over 1+z}\left(\sqrt{{\Omega_{hde}\over
  c^2}+\Omega_k}+{1\over2}-{3\Omega_{\Lambda}+\Omega_k \over 2(1-\Omega_{hde})}\right),
\end{equation}

where the fractional density of curvature and cosmological constant are:
\begin{equation}
  \Omega_{k} = {\rho_{k} \over 3M_p^2H^2} = {\Omega_{k0}(1 + z)^2 \over E^2},
\end{equation}
\begin{equation}
  \Omega_{\Lambda} = {\rho_{\Lambda} \over 3M_p^2H^2} = {\Omega_{\Lambda0} \over E^2}.
\end{equation}
Here the subscript '0' denotes the present-day value. Using the initial condition $E(z = 0) = 1$ and $\Omega_{hde0} = \Omega_{de0} - \Omega_{\Lambda0}$ where $\Omega_{hde0}$ and $\Omega_{de0}$ denote the present-day fractional density of HDE and total DE, we can solve Eq.(\ref{elhde}) and Eq.(\ref{Ohde}) numerically.

The observational constraints of the $\Lambda$HDE model has been briefly studied in \cite{ref_lhde}.
In a work in preparation \cite{ref_lhde1}, several cosmological observations, including type Ia Supernova, cosmic microwave background, baryon acoustic oscillation and growth factor, are used to constrain the $\Lambda$HDE model.
The best-fit results are $\Omega_{de0}=0.716$, $\Omega_{\Lambda0}=0.564$, $c=0.171$ and $\Omega_{k0}=-0.0002$.
As mentioned above, the main aim of this work is to distinguish the $\Lambda$HDE models with different $\Omega_{\Lambda0}$ or different $c$.
In the process of analysis, we use these best-fit values to set the other model parameters.
\vspace*{1mm}

\section{Diagnostic tools} \label{tools}

\subsection{The statefinder hierarchy} \label{statefinder}

Statefinder hierarchy \cite{ref_sttfd1,ref_sttfd2} is a powerful geometry diagnostic,
which makes use of the information from high-order derivatives of scale factor $a$ to distinguish different DE models from the $\Lambda$CDM model.
It has been already used to study various DE models \cite{ref_sttfd_model1, ref_sttfd_model2, ref_sttfd_model3}.

To derive the expression of statefinder hierarchy, first we Taylor-expand the the scale factor $a(t)/a_0$ around the present epoch $t_0$:
\begin{equation}
\frac{a(t)}{a_0} = 1 + \sum_{n=1}^\infty\frac{A_n(t_0)}{n!}[H_0(t-t_0)]^n
\end{equation}
where
\begin{equation}
  \label{An}
A_n = \frac{a(t)^{(n)}}{a(t)H^n}, n \in N
\end{equation}
with $a(t)^{(n)}=\frac{d^na(t)}{dt^n}$.
Notice that $A_2 = -q$ represents the deceleration parameter. For $\Lambda$CDM model, the functions above can be expressed by the fractional matter density $\Omega_m$:
\begin{equation}
  \label{Ans}
  \begin{split}
A_2 &= 1 - \frac{3}{2}\Omega_m,\\
A_3 &= 1,\\
A_4 &= 1 - \frac{9}{2}\Omega_m,...
  \end{split}
\end{equation}
Thus, we define the statefinder hierarchy $S_n$ as \cite{ref_sttfd2}
\begin{equation}
  \label{Sns}
  \begin{split}
S_2 &= A_2 + \frac{3}{2}\Omega_m,\\
S_3 &= A_3,\\
S_4 &= A_4 + \frac{9}{2}\Omega_m,...
  \end{split}
\end{equation}
it is obvious that every parameter of $S_n$ remains unity for $\Lambda$CDM model during cosmic evolution.
In \cite{ref_sttfd2}, the authors further introduce two statefinder hierarchy members: $S_3^{(1)}$ and $S_4^{(1)}$, which are given by
\begin{equation}
  \label{Sn1}
  \begin{split}
S_3^{(1)} &= A_3,\\
S_4^{(1)} &= A_4 + 3(1 + q).
  \end{split}
\end{equation}
In this work, we just use $S_3^{(1)}$ and $S_4^{(1)}$ to diagnose the $\Lambda$HDE model. From Eq.(\ref{elhde}) and Eq.(\ref{Ohde}), we can derive them for the $\Lambda$HDE case, which can be expressed as
\begin{equation}
  \label{S31}
S_3^{(1)} = -q' + q + 2q^2,
\end{equation}
\begin{equation}
  \label{S41}
S_4^{(1)} = -q'' + (3 + 7q)q' + (3 + q - 7q^2 - 6q^3),
\end{equation}
where the deceleration parameter takes the form
\begin{equation}
  \label{q}
q = \frac{1}{2}(1 + 3w_{hde}\Omega_{hde} -3\Omega_\Lambda - \Omega_k) = \frac{1}{2}(1 + 3w_{de}\Omega_{de} - \Omega_k).
\end{equation}
Note that the prime denotes the derivative with respect to $s = \mathrm{ln}a$.

\subsection{The fractional growth parameter} \label{growth}

In linear pertubation theory, the perturbation of the matter density $\rho_m$ is defined as $\delta_m = \delta\rho_m / \rho_m$.
It satisfies the equation \cite{ref_growtheq}
\begin{equation}
   \label{deltam}
\ddot\delta_m + 2 {\dot a \over a} \dot\delta_m - {4\pi \over M_p^2}\rho_m\delta_m = 0.
\end{equation}
Note that the dot above denotes the derivative with respect to time $t$.
The growth rate of linear density pertubation is defined as $f = d\mathrm{ln}\delta_m / d\mathrm{ln}a$. The one-order and two-order derivatives of $\delta_m$ with respect to time $t$ can be written as:
\begin{equation}
   \label{dotdeltam}
\dot\delta_m = f H \delta_m,
\end{equation}
\begin{equation}
   \label{ddotdeltam}
\ddot\delta_m = (\dot f H + f \dot H + f^2 H^2) \delta_m.
\end{equation}

Substituting Eqs. (\ref{dotdeltam}) and (\ref{ddotdeltam}) into Eq.(\ref{deltam}), we can derive the equation which determines $f$:
\begin{equation}
\label{f}
{df \over dz} = \frac{f^2 + 2f - f_1(z)}{1 + z} - \frac{dH / dz}{H}f,
\end{equation}
where $f_1(z) = \rho_m / H^2 = \frac{3\Omega_{m0}(1 + z)^3}{8\pi G E^2}$, and the present-day fractional matter density $\Omega_{m0} = 1 - \Omega_{de0} - \Omega_{k0}$. Using the initial condition $f(z = 0) = 1$, this equation can be numerically solved for the $\Lambda$CDM model and the $\Lambda$HDE model. Based on $f(z)$, another null diagnostic, which is called fractional growth parameter $\epsilon$, is defined as \cite{ref_growth1,ref_growth2}
\begin{equation}
\epsilon(z) = f(z) / f_{\mathrm{\Lambda CDM}}(z).
\end{equation}

In \cite{ref_sttfd2}, the author also introduced a quantity called composite null diagnostic (CND), which is a combination of statefinder hierarchy members and fractional growth parameter.
In this paper, we use two CND pairs \{$S_3^{(1)}, \epsilon$\} and \{$S_4^{(1)}, \epsilon$\} to diagnose the $\Lambda$HDE model.

\vspace*{1mm}

\section{Result} \label{result}

In this section, we use the three diagnostic tools to explore the impacts of various model parameters of $\Lambda$HDE.
To discuss the affects of key parameters $\Omega_{\Lambda0}$ and $c$, we vary $\Omega_{\Lambda0}$ and $c$ respectively, while fixing other parameters according to the best-fit values shown in Sec. \ref{model}.

%--------------Fig. 1
\begin{figure*}
\centering
\includegraphics[width = 7in]{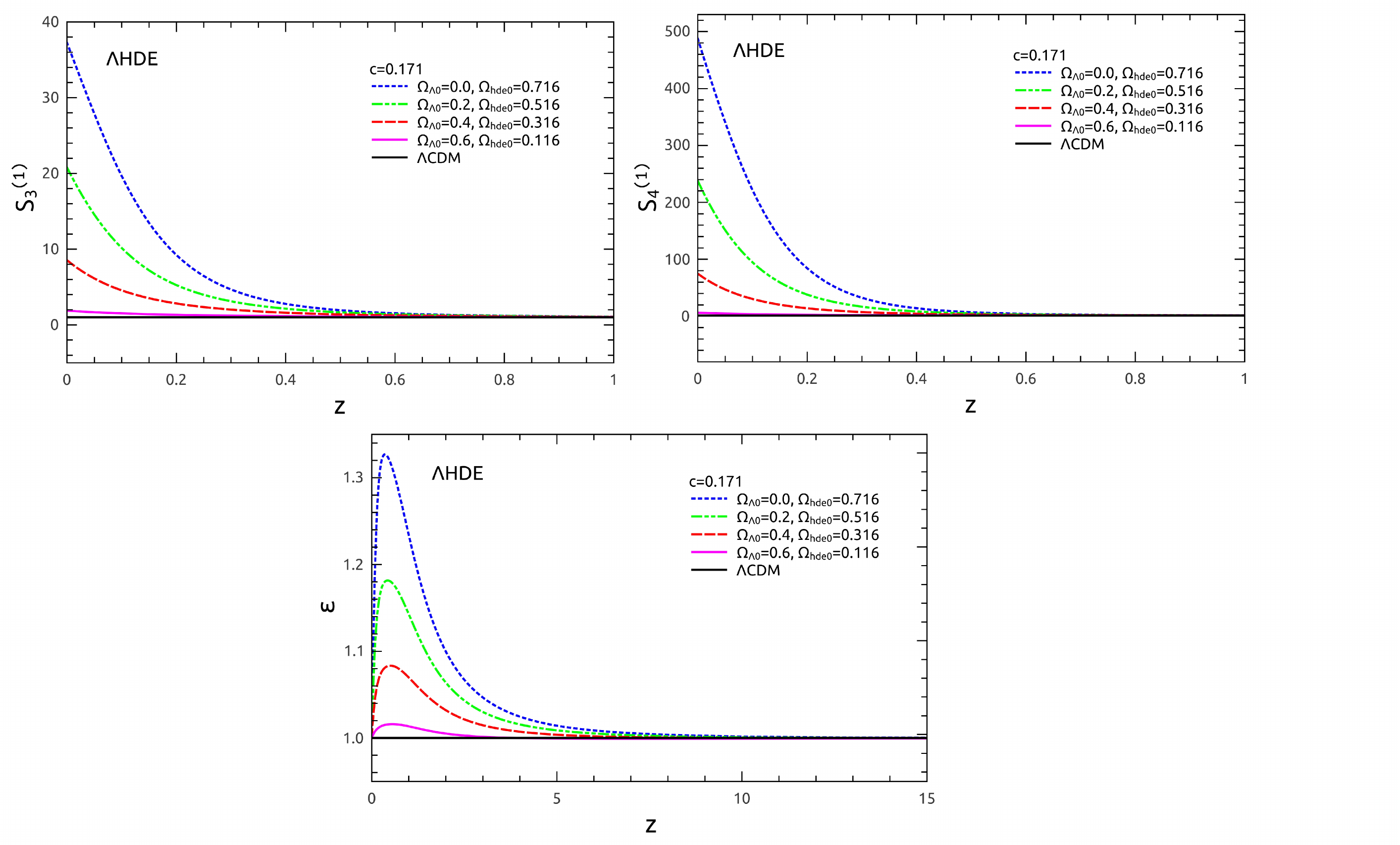}
\caption{\footnotesize{(colour online) The upper left, upper right and lower panels show the evolutionary trajectories of $S_3^{(1)}(z)$, $S_4^{(1)}(z)$ and $\epsilon(z)$ for the $\Lambda$HDE model respectively, with varying $\Omega_{\Lambda0}$ and other parameters fixed. Different linetypes correspond to different values of $\Omega_{\Lambda0}$. To make a comparison, we also show the result of the $\Lambda$CDM model in this figure as the solid horizontal lines.}}
\label{f_O}
\end{figure*}
Firstly, we study the impacts of $\Omega_{\Lambda0}$.
In Fig. \ref{f_O}, we plot the evolutionary trajectories of $S_3^{(1)}(z)$, $S_4^{(1)}(z)$ and $\epsilon(z)$ for the $\Lambda$HDE model, while fixing $c = 0.171$ and varying $\Omega_{\Lambda0}$ among 0, 0.2, 0.4 and 0.6.
In order to make a comparison, we also plot the result of the $\Lambda$CDM model as a solid line.
%------------------------------results of Fig. 1
For all the four cases associated with different $\Omega_{\Lambda0}$, all the curves of $S_3^{(1)}$, $S_4^{(1)}$ and $\epsilon$
 have similar evolutionary trajectories.
For example, all the curves of $S_3^{(1)}$ and $S_4^{(1)}$ have significant differences at low-redshift, and descend monotonically at higher-redshift.
In addition, all the curves of $\epsilon$ have convex vertice at $z \sim 0.4$ and descend monotonically at higher-redshift.
These results show that adopting different values of $\Omega_{\Lambda0}$ only has quantitative impacts on the cosmic evolution of the $\Lambda$HDE model.

%--------------Fig. 2
\begin{figure*}
\centering
\includegraphics[width = 7in]{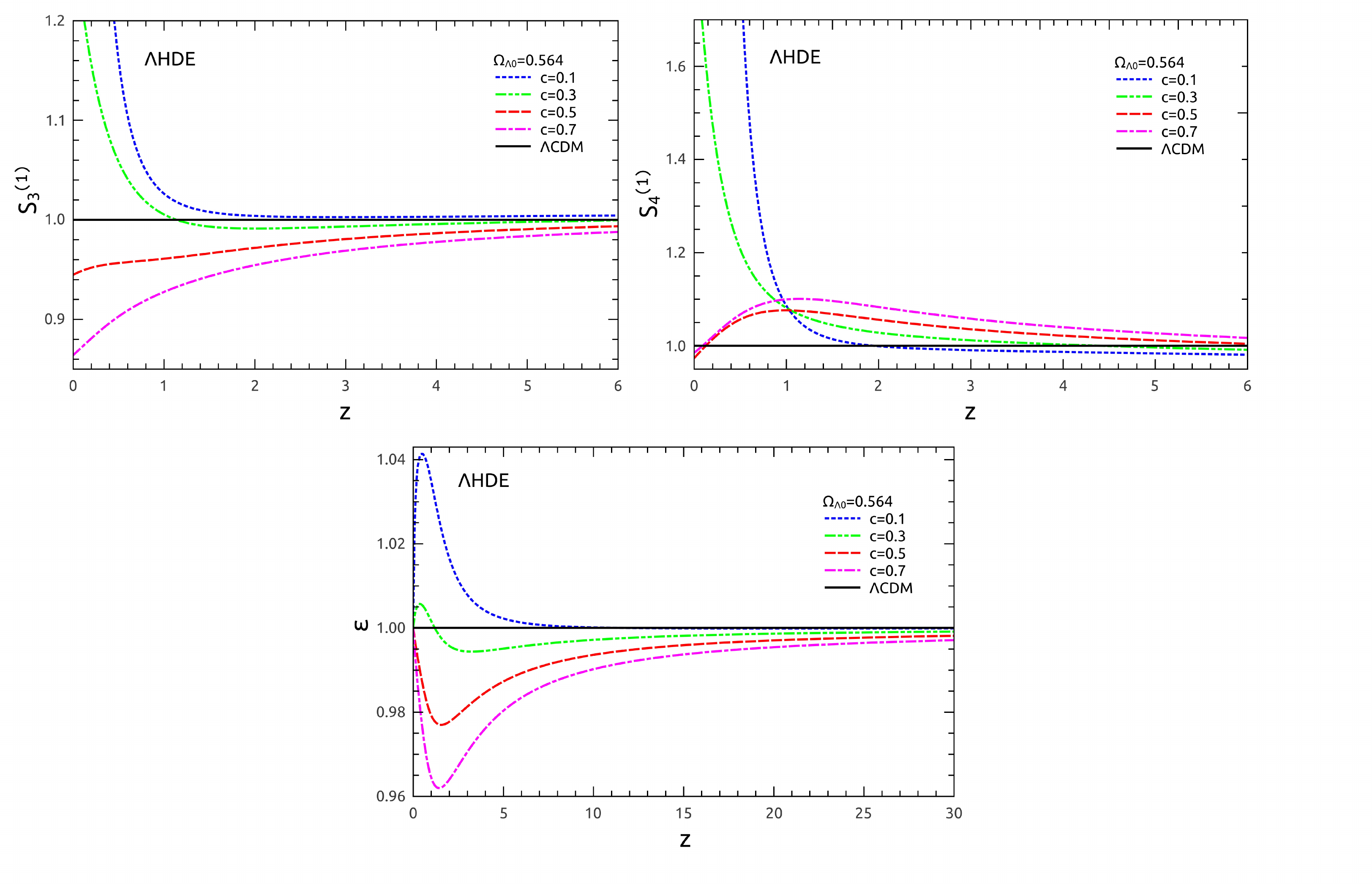}
\caption{\footnotesize{(colour online) The upper left, upper right and lower panels show the evolutionary trajectories of $S_3^{(1)}(z)$, $S_4^{(1)}(z)$ and $\epsilon(z)$ for the $\Lambda$HDE model respectively, with varying $c$ and other parameters fixed. Different linetypes correspond to different values of $c$.
To make a comparison, we also show the result of the $\Lambda$CDM model in this figure as the solid horizontal lines.}}
\label{f_c}
\end{figure*}
Then we study the impacts of $c$.
In Fig. \ref{f_c}, we plot the evolutionary trajectories of $S_3^{(1)}(z)$, $S_4^{(1)}(z)$ and $\epsilon(z)$ for the $\Lambda$HDE model, while fixing $\Omega_{\Lambda0} = 0.564$ and varying $c$ among 0.1, 0.3, 0.5 and 0.7.
To make a comparison, we also plot the result of the $\Lambda$CDM model as a solid line.
%------------------------------results of Fig. 2
Different from the cases of $\Omega_{\Lambda0}$, the curves of $S_3^{(1)}$, $S_4^{(1)}$ and $\epsilon$ associated with different $c$ have different evolutionary trajectories.
For example, all the curves of $S_3^{(1)}$ and $S_4^{(1)}$ have different evolutionary behavior at low-redshift, and have a trend of coincidence at high-redshift.
In addition, among the curves of $\epsilon$, the curve associated with $c = 0.1$ has a convex vertex above the $\Lambda$CDM line;
the curves associated with $c = 0.5$ and $c = 0.7$ have concave vertice under the $\Lambda$CDM line;
while the curve associated with $c = 0.3$ has both two kinds of vertice.
These results show that adopting different values of $c$ has qualitative impacts on the cosmic evolution of the $\Lambda$HDE model.

%------------Tab. 1, 2
\begin{table}[H]

  \caption{\footnotesize{The present values of $S_{30}^{(1)}$ and $S_{40}^{(1)}$, and their differences $\Delta S_{30}^{(1)} = S_{30}^{(1)}(\text{max}) - S_{30}^{(1)}(\text{min})$, $\Delta S_{40}^{(1)} = S_{40}^{(1)}(\text{max}) - S_{40}^{(1)}(\text{min})$. A varying $\Omega_{\Lambda0}$ is adopted in the analysis.}}
  \label{t_O}
\begin{center}\footnotesize \doublerulesep 0.2pt
\begin{tabular}{ccccc}
\hline\hline
$\Omega_{\Lambda0}$&~~0&~~0.2&~~0.4&~~0.6\\
\hline
c&~~&~~~~~~~~~~~~~~~~~0.171\\
\hline
$S_{30}^{(1)}$&~~37.2915&~~20.8119&~~8.5451&~~1.8763\\
$S_{40}^{(1)}$&~~487.4201&~~237.5444&~~74.5956&~~6.0481\\
\hline
$\Delta S_{30}^{(1)}$&~~&~~~~~~~~~~~~~~~~~35.4155\\
$\Delta S_{40}^{(1)}$&~~&~~~~~~~~~~~~~~~~~481.3720\\
\hline\hline
\end{tabular}
\end{center}
\end{table}
\begin{table}[H]
  \caption{\footnotesize{The present values of $S_{30}^{(1)}$, $S_{40}^{(1)}$, and their differences $\Delta S_{30}^{(1)} = S_{30}^{(1)}(\text{max}) - S_{30}^{(1)}(\text{min})$, $\Delta S_{40}^{(1)} = S_{40}^{(1)}(\text{max}) - S_{40}^{(1)}(\text{min})$. A varying $c$ is adopted in the analysis.}}
  \label{t_c}
\begin{center}\footnotesize \doublerulesep 0.2pt
\begin{tabular}{ccccc}
\hline\hline
$\Omega_{\Lambda0}$&~~&~~~~~~~~~~~~~~~~~0.564\\
\hline
c&~~0.1&~~0.3&~~0.5&~~0.7\\
\hline
$S_{30}^{(1)}$&~~6.5876&~~1.3002&~~0.9447&~~0.8640\\
$S_{40}^{(1)}$&~~72.3008&~~1.9622&~~0.9734&~~0.9850\\
\hline
$\Delta S_{30}^{(1)}$&~~&~~~~~~~~~~~~~~~~~5.7236\\
$\Delta S_{40}^{(1)}$&~~&~~~~~~~~~~~~~~~~~71.3158\\
\hline\hline
\end{tabular}
\end{center}
\end{table}
To compare $S_3^{(1)}$ and $S_4^{(1)}$ for the cases associated with different $\Omega_{\Lambda0}$ or different $c$ with more details, we list the present values of $S_3^{(1)}$ and $S_4^{(1)}$ and their differences in tables \ref{t_O} and \ref{t_c}.
Notice that $S_{30}^{(1)}$ and $S_{40}^{(1)}$ are the present values of $S_3^{(1)}$ and $S_4^{(1)}$, while
$\Delta S_{30}^{(1)} = S_{30}^{(1)}(\text{max}) - S_{30}^{(1)}(\text{min})$ and $\Delta S_{40}^{(1)} = S_{40}^{(1)}(\text{max}) - S_{40}^{(1)}(\text{min})$ denotes the differences between the maximum and minimum values of $S_{30}^{(1)}$ and $S_{40}^{(1)}$.
%------------------------results of Tab. 1, 2
From these two tables, we can see that $\Delta S_{40}^{(1)}$ is much larger than $\Delta S_{30}^{(1)}$.
This means that compared with $S_3^{(1)}$, $S_4^{(1)}$ can give larger differences among the cosmic evolutions of the $\Lambda$HDE model associated with different $\Omega_{\Lambda0}$ or different $c$.
%In the cases of different $c$, when $c = 0.4$, 0.6 and 0.8, both $S_{30}^{(1)}$ and $S_{40}^{(1)}$ are close to 1; when $c = 0.2$, $S_{30}^{(1)} = 2.4243$ and $S_{40}^{(1)} = 9.4428$ are remarkably larger than 1.

%--------------Fig. 3
\begin{figure*}
\centering
\includegraphics[width = 7in]{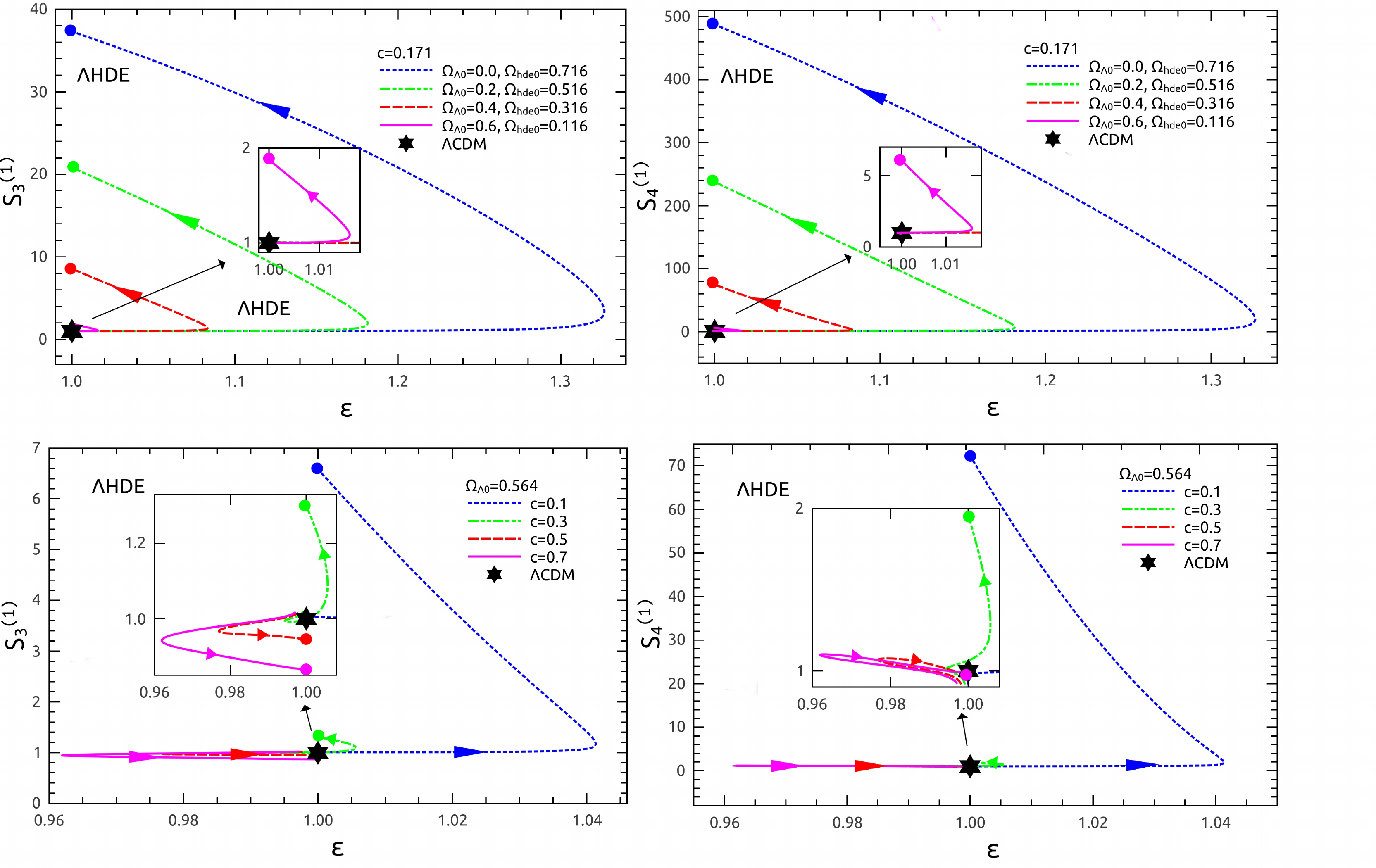}
\caption{\footnotesize{(colour online) The evolutionary trajectories of the CND pairs \{$S_3^{(1)}, \epsilon$\} (left panels) and \{$S_4^{(1)}, \epsilon$\} (right panels) for different $\Omega_{\Lambda0}$ (upper panels) and different $c$ (lower panels).
The current values of \{$S_3^{(1)}, \epsilon$\} and \{$S_4^{(1)}, \epsilon$\} for the $\Lambda$HDE models are marked by the round dots, and the arrows indicate the time directions of cosmic evolution, i.e. $z \to 0$.
To make a comparison, we also show the result of the $\Lambda$CDM model in this figure as star-shape points.
}}
\label{f_es3s4}
\end{figure*}
As can be seen in Figs. \ref{f_O} and \ref{f_c}, compared with $S_3^{(1)}$ and $S_4^{(1)}$, $\epsilon$ has very different evolutionary trajectories.
So we can use a composite null diagnostic (CND), which is a combination of \{$S_3^{(1)}, S_4^{(1)}$\} and $\epsilon$, to diagnose the $\Lambda$HDE model.
The results are shown in Fig. \ref{f_es3s4}, which gives the evolutionary trajectories of the CND pairs \{$S_3^{(1)}, \epsilon$\} (left panels) and \{$S_4^{(1)}, \epsilon$\} (right panels) for different $\Omega_{\Lambda0}$ (upper panels) and different $c$ (lower panels).
The current values of \{$S_3^{(1)}, \epsilon$\} and \{$S_4^{(1)}, \epsilon$\} are marked by the round dots, and the arrows indicate the time directions of evolution, i.e. $z \to 0$.
To make a comparison, we also plot the result of $\Lambda$CDM as star-shape points.
%--------------------------------result of Fig. 3
For the cases of varying $\Omega_{\Lambda0}$, the curves of CND pairs only have quantitative differences:
at high-redshift region, each curve of \{$S_3^{(1)}, \epsilon$\} and \{$S_4^{(1)}, \epsilon$\} starts from the neighbourhood of the star symbol of the $\Lambda$CDM model, then evolves towards the direction of the increase of $\epsilon$; after passing a turning point, it continues evolving towards the direction of the decrease of $\epsilon$ and the increase of \{$S_3^{(1)}, S_4^{(1)}$\}.
On the other side, for the cases of varying $c$, the curves of CND pairs have qualitative differences:
the curve associated with $c = 0.1$ has a similar evolutionary trajectory with the case of varying $\Omega_{\Lambda0}$;
in contrast, the curve associated with $c = 0.5$ or $c = 0.7$ evolves towards an opposite direction, and returns to the neighbourhood of the star symbol of the $\Lambda$CDM model.
These results further verify the conclusions of Figs. \ref{f_O} and \ref{f_c}.
Moreover, since using a single diagnostic tool can only give 1-dimensional evolution information, adopting CND has much stronger ability to diagnose the $\Lambda$HDE model.

\section{Conclusion and Discussion} \label{summary}

%================motivation
In a latest paper \cite{ref_lhde}, a new DE model called $\Lambda$HDE was proposed.
In this model, DE consists of two parts: cosmological constant $\Lambda$ and HDE.
Two key parameters of this model are the fractional density of cosmological constant $\Omega_{\Lambda0}$, and the dimensionless HDE parameter $c$.
Since these two parameters determine the dynamical properties of DE and the destiny of universe, it is important to study the impacts of different values of $\Omega_{\Lambda0}$ and $c$ on the $\Lambda$HDE model.

%=================tools
Two diagnostic tools are often used to analyze various DE models.
One is statefinder hierarchy \{$S_3^{(1)}, S_4^{(1)}$\},
another is the fractional growth parameter $\epsilon$.
In addition, the CND pair, which is a combination of these two quantities, is also widely used to diagnose DE models.
Therefore, the main aim of this work is making use of these diagnostic tools to distinguish the $\Lambda$HDE models with different $\Omega_{\Lambda0}$ or different $c$.

%=====================result
The conclusions of this work are as follows:
Firstly, from Figs. \ref{f_O} and \ref{f_c}, we find that adopting different values of $\Omega_{\Lambda0}$ only has quantitative impacts on the evolution of the $\Lambda$HDE model, while adopting different $c$ has qualitative impacts;
Secondly, by comparing tables \ref{t_O} and \ref{t_c}, we find that compared with $S_3^{(1)}$, $S_4^{(1)}$ can give larger differences among the cosmic evolutions of the $\Lambda$HDE model associated with different $\Omega_{\Lambda0}$ or different $c$;
Thirdly, by analyzing Fig. \ref{f_es3s4} in details, we find that compared with the case of using a single diagnostic, adopting a CND pair has much stronger ability to diagnose the $\Lambda$HDE model.

%====================future
In this work, only two kinds of diagnostic tools, statefinder hierarchy and fractional growth parameter, are used to diagnose the $\Lambda$HDE model.
There are some other diagnostic tools, such as $w-w'$ analysis \cite{ref_wdw1, ref_wdw2}.
It is interesting to make use of these diagnostic tools to analyze various DE models, and compare the advantages and disadvantages of various diagnostic tools.
In addition, it is also very interesting to study
the impacts of supernova's systematic uncertainties  \cite{ref_beta1,ref_beta2,ref_beta3,ref_beta4,ref_beta5,ref_beta6} on the $\Lambda$HDE model.
These will be done in future works.

\vspace*{2mm} \Acknowledgements{\bahao We wish to thank Li Nan and Cheng Cheng for valuable suggestions and technological supports.
SW is supported by the National Natural Science Foundation of China under Grant No. 11405024.}

\end{multicols}

\end{document}